\documentclass[useAMS,usenatbib]{mn2e}
\usepackage {graphicx}
\usepackage{natbib}
\usepackage{aas_journals}
\usepackage{times}
\usepackage{color}

\title[Cross-correlation of WMAP7 and the WISE Full Data Release]{Cross-correlation of WMAP7 and the WISE Full Data Release}
\author[A. Kov\'acs et al. ]{Andr\'as Kov\'acs$^{1,4}$, Istv\'an Szapudi$^{2}$, Benjamin R. Granett$^{3}$ and Zsolt Frei$^{1,4}$ {}\\
$^{1}$Department of Atomic Physics, E\"otv\"os Lor\'and University, 1117 P\'azm\'any P\'eter s\'et\'any 1/A Budapest, Hungary\\
$^{2}$Institute for Astronomy, University of Hawaii 2680 Woodlawn Drive, Honolulu, HI, 96822, USA\\
$^{3}$Istituto Nazionale di Astrofisica - Osservatorio Astronomico di Brera, Via E. Bianchi 46, 23807 Merate, Italy\\
$^{4}$ MTA-ELTE EIRSA "Lendulet" Astrophysics Research Group}

\begin{document}

\date{Accepted 2013 January 02}

\pagerange{\pageref{firstpage}--\pageref{lastpage}} \pubyear{2012}

\maketitle

\label{firstpage}
\begin{abstract}
We measured the cross-correlation of the Wilkinson Microwave Anisotropy Probe (WMAP) 7 year temperature map and the full sky data release of the Wide-field Infrared Survey Explorer (WISE) galaxy map. Using careful mapmaking and masking techniques we find a positive cross-correlation signal. The results are fully consistent with a $\Lambda$CDM Universe, although not statistically significant. Our findings are robust against changing the galactic latitude cut from $|b|>10$ to $|b|>20$ and no color dependence was detected when we used WMAP Q, V or W maps. We confirm higher significance correlations found in the preliminary data release. The change in significance is consistent with cosmic variance.
\end{abstract}


\section{Introduction}
Cosmological supernovae measurements and Cosmic Microwave Background (CMB) fluctuations support cosmological models in which the cosmic energy density is dominated by Dark Energy (DE) at the present epoch \citep{wmap,RiessEtal98}. In such theories the current accelerating expansion and the decay of gravitational potentials are predicted. Therefore, the presence of DE is manifested in both geometrical and dynamical forms.

Dark Energy comes to dominate the energy density at late times, $z<2$, and so the primordial fluctuations in the CMB alone do not provide a sensitive probe.  However, DE may leave a signal in the secondary anisotropies that are imprinted on the microwave background radiation.  The Integrated Sachs-Wolfe effect (ISW) \citep{b17,b18} is an example of a secondary anisotropy:  CMB photons passing through a changing graviational potential become slightly hotter or colder. In a flat and matter-dominated Universe the potential is constant on large scales thus gravitational blueshifts and redshifts cancel along the photon path. However, in a Universe dominated by DE there is a net energy difference between entering and leaving a potential well due to the decay.  Thus, the detection of the linear ISW effect provides direct evidence for dark energy in the $\Lambda$CDM model. Furthermore, alternative gravity models provide predictions for the ISW effect and may be directly tested with ISW observations \citep{gianEtal2010}.

The ISW signal may be detected through cross-correlation of large-scale structure surveys with the CMB temperature maps. The correlation is weak, generally less than a 1 $\mu$K signal is expected, orders of magnitude below the primary fluctuations. Futhermore, the ISW effect is strongest on large angular scales where cosmic variance is also large, making the measurement even more cumbersome.

Several measurements have been performed to uncover the ISW signal: positive cross-correlations were measured using galaxy data from the Sloan Digital Sky Survey (SDSS) and WMAP \citep{b3,b4,b5,GranettEtal09,b6}. Other successful attempts were \cite{b7} based on APM galaxies, \cite{b9,b10} using radio data and \cite{b11,b12} in which the hard X-ray background was investigated. Besides, \cite{b8,b13} and \cite{b14} used infrared galaxy samples to characterize the ISW signal. The typical ISW significance in the papers above is around 2-3$\sigma$. Comprehensive studies using combinations of data sets were carried out by \cite{ho} and \cite{gianEtal08,gian}.

The Wide-field Infrared Survey Explorer (WISE) all-sky survey is an attractive dataset for ISW studies. The survey effectively probes low redshift $z<0.3$ with a high source density.  Using the preliminary data release (PDR) covering 10000 sqr deg. \cite{tomo} cross-correlated a WISE galaxy sample with the Cosmic Microwave Background, finding a $3\sigma$ detection, although with three times the amplitude expected in $\Lambda$CDM. In this paper, we rexamine this finding  using the full-sky data release (FDR) of the WISE survey and the WMAP 7-year dataset.

The structure of this paper is as follows. In Section 2 we describe the data we used in particular. Section 3 describes our methods including the theoretical expectations, simulations and measurements. Finally, in Section 4 the statistical significances are presented and systematic effects are discussed. 

\section{CMB data and Galaxy map}
We used the best achieveable versions of the CMB data products and focused on the reliability of our new galaxy sample. In this section we describe our map-making and masking procedures.
\subsection{CMB map}
The 7-year WMAP temperature data were downloaded from the LAMBDA website \footnote{\texttt{http://lambda.gsfc.nasa.gov/}} \citep{wmap}. WMAP data is affected by noise and contaminations both from point sources and the Milky Way. Among all, the Q, V and W maps have the least galactic contamination. We used the foreground reduced version of these maps and the CMB Extended Temperature Mask was chosen to avoid contaminations.

Using HEALPIX \citep{b20}, NSIDE=128 repixelized versions of the maps were created. Galactic foregrounds and known point sources are quite surely excluded and 71\% of the sky is unmasked.
\begin{figure}
\begin{center}
\includegraphics[width=80mm]{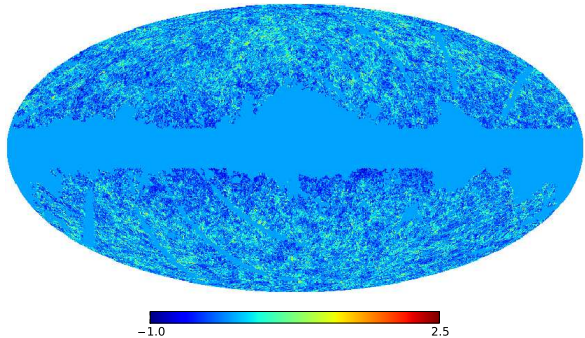}
\caption{WISE allsky galaxy sample, together with our mask including the stripes and WMAP's mask area.}
\label{allsky}
\end{center}
\end{figure}
\subsection{WISE galaxy map}
The density map of the galaxies was prepared using the full data release of the WISE project \citep{wise}. The WISE satellite surveyed the sky at four different wavelengths: 3.4, 4.6, 12 and 22 $\mu$m. We used different bands to separate stars from galaxies using color-color plots. Following \cite{tomo} we select sources to a flux limit of W1 $\leq $ 15.2 mag to have a uniform dataset.

According to \cite{tomo} the majority of stars near to the galactic plane have a $W_{3.4}-W_{4.6}$ $\leq $ 0.2 mag color. Moreover, it was found that a $W_{4.6}-W_{12}$ $\leq $ 2.9 mag selection reduced the stellar contamination. We confirm these findings in the FDR and followed the same procedure for star-galaxy separation.

Our galaxy sample exhibits stripe-like overdensities on the map in several locations. While \cite{tomo} applied handmade cutouts in their mask to exclude regions with unusually high number counts, we understand that the stripe-like features originated from the observational strategy of WISE and the position of the moon. We realized that the moon-contamination flag may be used to properly mask these regions. We forced out pixels in which the 'moonlev' flag is higher than 3 in at least one of the bands. This means that the fraction of the used image frames suffered by moon-contamination is higher than 30\%. Masking regions based upon the moon contamination flag effectively removes the stripe pattern. We cannot address any further residual effects of the moon contamination outside our mask area.
We have found that with a more conservative magnitude limit of W1 $<$ 14.9 overdensities along stripes are reduced in width but do not disappear.

\begin{figure}
\begin{center}
\centering
\includegraphics[width=90mm]{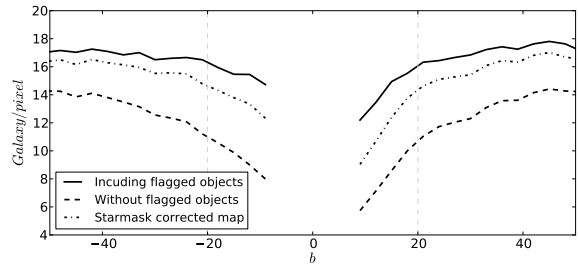}
\caption{Gradients in galaxy density can be detected as a function of galactic latitude $b$. The effect is less significant if objects with warning flags are included in the sample, since the density drop is $\approx$ 4\% at $|b|=20$ compared to its value at $|b|=50$.}
\label{bgrad}
\end{center}
\end{figure}

\begin{table} \centering \caption{Mask properties of the WISE galaxy sample. The first 3 rows correspond to allsky data with $b$ galactic latitude cuts. The last mask was made using only the preliminary survey area and an additional $|b|>10$ cut.} \label{symbols}
\begin{tabular}{@{}lcccccc} \hline Mask
& Area[$deg^2$] & $N_{gal}^{no flag}$ & $N_{gal}^{starcorr}$ & $N_{gal}$\\
\hline $|b|>10$ & 26443 & 1662995 & 2016563 & 2234370 \\ $|b|>15$   & 25248 & 1617745 & 1948454 & 2129013 \\ $|b|>20$  & 23167 & 1521755 & 1814156 & 1945517  \\ \hline $|b|>10$ & 10967 & 782502 & 976209 & 1057073 \end{tabular}
\end{table}

Sources in WISE may also be contaminated by artefacts including the halos of bright stars, ghost images and diffraction spikes and special data-quality flags exist to handle these problems. If the dataset is filtered using these flags we possibly lose real galaxies. However, if we do not use flags to create a conservative catalog then stars or galaxies with insufficient parameters can appear in the data. \cite{tomo} used additional pixels in their mask to exclude regions where the abundance of these unreliable objects is high, but did not filter the whole galaxy sample. We investigated both cases and we have found this choice is important especially on large scales.

We find a gradient in the galaxy density with Galactic latitude with fewer galaxies near the Galactic plane.
An empirical correction was developed by \cite{tomo} in which a mean correction is computed in galactic latitude bins to artificially `flatten' the distribution at $|b|<20$.

We attribute the gradient to stars near the Galactic plane masking background galaxies.  The problem is made more severe due to the broad point spread function of WISE (6-12 arcsec). We use the Tycho2 star catalogue which reaches a depth of $V<13$ mag \citep{tycho} to measure the survey area lost around each star.  We calibrated a mean relation between $V$ magnitude and star halo radius $R$ for WISE $R=9.52 - 0.74 V$ for $R$ in arcminutes.
Any detected sources within this radius of a Tycho star is removed.  We then construct a map of the lost area by summing the area attributed to stars in each Healpix pixel. This map is then used to normalise the galaxy counts. Figure \ref{bgrad} shows the result of this correction.

Apparently, the density gradient does not disappear, so only a modest correction is possible with our method. Interestingly, we find that the gradient is higher when flagged sources are not included in the sample. The exact source of the gradient is unexplored, but we show that this effect did not mess up our measurements, our findings are robust.

As described, regions nearby the plane of the Milky Way are potentially contaminated, and we consider the most appropriate solution to use $|b|>20$ regions. To perform tests with only the preliminary sky coverage area a mask was created using the area covered by the preliminary survey.

\subsection{Redshift distribution}
In order to calculate a theoretical ISW expectation, redshift information is needed. Since WISE is a photometric survey without spectroscopy, the selected galaxies were cross-identified with sources from GAMA (Galaxy and Mass Assembly, \cite{gama}) sample that has spectroscopic redshift of $\sim$ 200000 galaxies. Using the overlapping part of the two surveys we have found a pair for 82\% of the galaxies with a $3"$ matching radius. We estimated an accidental matching rate of 0.1\% for this analysis using random points with WISE density. The matched sample has a median redshift $\bar{z} \approx$ 0.15. The obtained approximate redshift distribution provides a basis to calculate a theoretical cross-power spectrum in this redshift range.

\section{Results}
In this section we discuss the results using the CMB and galaxy datasets. We also elaborate on the most important theoretical and simulated considerations related to ISW detection and further analysis.

\subsection{WISE-WMAP cross-correlation}
We calculate the cross power spectrum using a fast quadratic estimator SpICE (Spatially Inhomogenous Correlation Estimator, \cite{Szapudi,SzapudiA,SzapudiB}). The individual band powers are binned logarithmically, the boundaries are $l=6,8,11,16,22,31,44,61$ and $87$ this means the first band stands for $l=6,7$ etc. With this choice we avoid the lowest $l$ range, where cosmic variance is meaningful and it is easier to compare the results to \cite{tomo}, who used the same bins. Our measurement is introduced on Figure \ref{pspec}.

\subsection{Theory}
We derive the expected correlations and galaxy bias using WMAP7 best-fit $\Lambda CDM$ cosmological parameters \citep{wmap}. Following \cite{b14} a linear bias relation is considered to couple galaxy and matter overdenisities, $\delta_g = b\delta_m$. The two-dimensional projection of the 3D galaxy auto-correlation is given by
\begin{equation}
C_{gg}(l) =  b_{g}^{2} \frac{2}{\pi} \int dk \, k^2 \, P(k) {\biggl\vert \int dr \, r^2 \, \phi(r) \, j_l(kr) \biggl\vert }^2
\end{equation}
where $\phi(r) \propto \frac{dN(r)}{dz} \frac{dz}{dV}$ is a comoving coordinate with a normalization relation $\int\phi(r)r^2 dr=1$, and $j_l$ is a spherical Bessel function.
Independent determination of $b_g$ is not possible in linear theory, because $\sigma_8$ acts to renormalize the power spectrum, $C_l \propto {(b\sigma_8)}^2$. Thus we fit only for $b_g$ and keep $\sigma_8$ fixed. CosmoPy \footnote{\texttt{http://www.ifa.hawaii.edu/cosmopy/}} and CAMB \footnote{\texttt{http://camb.info/}} were used to generate nonlinear matter power spectra with Halofit \citep{smith} at the median redshift of our galaxy sample.

We measure the galaxy-galaxy power spectrum with SpICE. The measurement is affected by Poissonian shot noise that has a form of $1/N$, where $N$ is the mean galaxy count per steradian. Actually, the impact is negligible, less than $10 \%$ at the maximum $l$ we used and less significant at larger scales where we expect to measure ISW. However, to be precise we substracted the noise and the amplitude of the theoretical model curve was fitted on the $6<l<100$ interval, i.e. angular scales down to $\sim$$2^{\circ}$. Our result is $b_g = b\sigma_8 = 1.04 \pm 0.05$.

To bring the $b_g$ parameter into use consider now the expression of the theoretical ISW signal. The cross-spectrum of a galaxy map and the CMB is given by
\begin{eqnarray}
{C_l}^{gT} = {b_g} \frac{6 \cdot T_{CMB}\Omega_m {H_0}^2 } {\pi c^2} \int dk \, k^2 \, P_k \cdot \nonumber \\ \int dr \, j_l(kr) k^{-2}\, \frac{d(1+z)D_1(z)}{dr} \int dr' \, j_l(kr') \, \phi(r')r'^{2}
\label{eqgT}
\end{eqnarray}
where $D_1(z)$ is the linear growth factor, the numerical result of this expression depends on the cosmology \citep{cooray}. 

\subsection{Simulations}
\begin{figure}
\includegraphics[width=90mm]{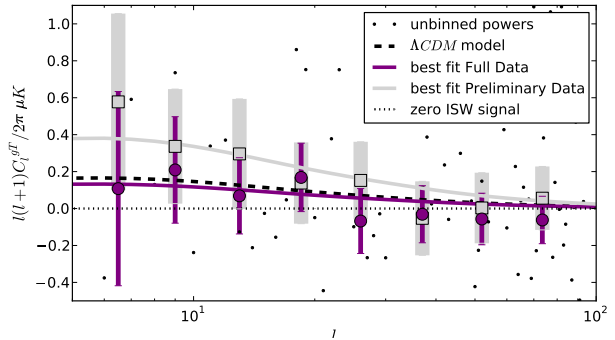}
\caption{Cross-correlation power spectra of WMAP7-WISE datasets using our $|b|>20$ mask, together with theoretical expectations for $\Lambda CDM$ cosmology and zero ISW detection. Magenta errorbars were computed using 1000 simulations and the diagonal elements of the covariance matrix. The light gray errorbars represent the variance of the ${C_l}^{FDR}-{C_l}^{PDR}$ differences in each bin to check if the change is consistent with cosmic variance.}
\label{pspec}
\end{figure}

\begin{figure}
\begin{center}
\includegraphics[width=70mm]{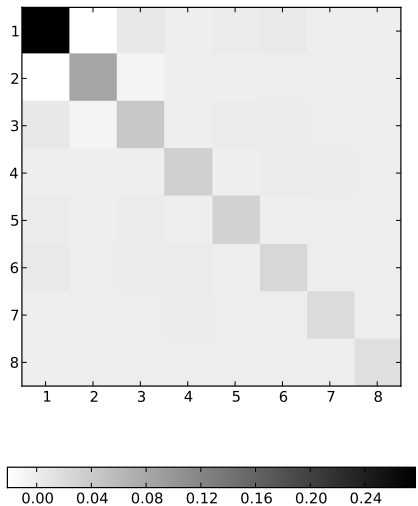}
\caption{Covariance matrix of the logarithmic spectral bins from 1000 random CMB simulations. Neighboring bands are not correlated in a good approximation.}
\label{covmtx}
\end{center}
\end{figure}

We simulated 1000 random CMB skies with $\Lambda CDM$ cosmological parameters using Healpix \texttt{synfast} to cross-correlate with our WISE galaxy density map. The power spectrum was calculated and binned into the 8 spectral bins given above. The covariance matrix estimated from these measurements is shown in Figure \ref{covmtx}. Neighboring bins are anti-correlated typically by 10\%. The diagonal elements were used to calculate the errorsbars shown on Figure \ref{pspec}.

\section{Significance tests}
\label{sign}
Again we follow \cite{b14}, now to determine the significance of our ISW detection. Consistency of our results was investigated with three hypotheses: null-detection of ISW, regular $\Lambda CDM$ model predictions and finally with a best-fit theoretical curve. Our statistics is based on the amplitude fit, we set the amplitude of the $\Lambda CDM$ theoretical curve to 1.0 and to 0.0 in the zero ISW case.
\subsection{Statistical tools}
We evaluate a ${\chi}^2$ statistic for each hypothesis which is the following:
\begin{equation}
{\chi}^2 = \sum_{ij} d_i C_{ij}^{-1} d_j
\end{equation}
where $C$ is the covariance matrix and $d_i=(C_{data}^{gT}-C_{hypo}^{gT})$.
$C_{hypo}^{gT}$ can be given by Equation \ref{eqgT} assuming various models or it is zero in the null-ISW case. Index $i$ labels the bins we use in the cross-spectrum.
Moving forward one step we define the likelihood of a hypothesis below:
\begin{equation}
\mathcal{L} \propto |C|^{-N/2} e^{-\frac{1}{2}{\bf d}^T C^{-1} {\bf d}}
\end{equation}
where N is the number of data points and with {\bf d} vector that was constructed using $d_i$ and $C$ matrix.

In fact, our tool to describe the detailed statistical properties of our test is $-2 \,ln\, (\mathcal{L}_1/\mathcal{L}_2) = \Delta \chi^{2}$ where the ratio of the two likelihoods is taken in a case of two different hypotheses and $\Delta \chi^{2}$ is calculated. In general, $\Delta \chi^{2} > 3$ is a strong evidence for a significant difference.

\begin{table} \centering \caption{Significance properties of our results are shown. $\chi^{2}$ values in the table are less than theoretically expected, this indicates that the error bars are possibly overestimated. The same was reported by Francis \& Peacock and Rassat et al. in their work. We performed Monte Carlo runs with 1000 and 3000 trials, but the covariance was robust. Moreover, the analytic estimate of the covariances and errors \citep{cabre_etal} agrees to 5\% with the MC outcomes.} \label{symbols}
\begin{tabular}{@{}lcccccc} \hline Mask & ISW Model & $\chi^{2}$& $\Delta \chi^{2}$ & Amplitude & S/N\\
\hline  & Zero & 3.07 & - & \\ $|b|>10$ &  Best-fit & 2.20 & 0.87 & $0.8\pm 0.9$ & 0.9 \\ &  $\Lambda CDM$ & 2.26 & 0.81 \\
\hline  & Zero & 2.71 &  - & \\ $|b|>15$ & Best-fit & 2.13 & 0.58 & $0.7\pm 0.9$ & 0.9 \\ &  $\Lambda CDM$ & 2.27 & 0.44  \\
\hline  & Zero & 2.32 & - & \\ $|b|>20$ & Best-fit & 1.63 & 0.69 & $0.8\pm 0.8$ & 1.0 \\ & $\Lambda CDM$ & 1.74 & 0.58  \\
\hline $|b|>10$ & Zero & 5.64 & - &  \\  preliminary & Best-fit & 2.91 & 2.73 & $2.3\pm 1.2$ & 1.9 &  \\ area only & $\Lambda CDM$ & 3.74 & 1.90 \\
\end{tabular}
\label{t2}
\end{table}
\subsection{Systematic effects}
Although we have taken care in source selection and masking we must address any residual systematic effects that may exist. Galactic dust or emission detected by WMAP could contribute to a correlation with WISE due to dust attenuation or the gradient in source density measured with Galactic latitude.  However, we performed all the significance tests using Q, V and W foreground reduced CMB maps and the maximum relative difference in a given spectral bin was 1.2\%. This fact lead us to the conclusion that there is no significant color dependence and effects from Galactic dust or emission must be minor.
We also checked the results using the WMAP team's Temperature Mask or Extended Temperature Mask but no meaningful difference was found.

On the other hand, we investigated several systematic effects related to our galaxy sample. Firstly, we analyzed different galactic latitude cuts. Our finding is robust, even if not significant, the results are summarized in Table \ref{t2}.

Next we considered the differences in the detection significance due to initial mapmaking. Our tests were applied to a map without flagged objects and it was obtained that the amplitude varies between 0.5 and 1.0 regardless of the starmask correction technique.

We repeated our analysis with W1 $\leq $ 14.9 mag limit to test the effects of faint sources. With this sample originating from a slightly different redshift distribution we found an increment in the ISW signal, but the errorbars were also high so the significance remained $\sim$1.0$\sigma$.

We also extended our cross-correlation analysis to $2 \leq l \leq 5$ multipoles and very weak positive cross-correlation was measured with extremely high errorbars. The results increased the significance only slightly, however, they were sensitive to the galactic cut.

In the light of the robustness of our results against different galactic cuts we argue that the stellar contamination is low or at least uniform in our galaxy sample and it does not affect the measurements. The upper limit is 18\% from the WISE-GAMA matching but a similar deficit of optical pairs was also reported using SDSS and WISE \citep{yanetal2012}. These facts all indicate that the different selection criteria of the infrared and the optical bands are responsible for the missing counterparts, rather than confusion with stars. In summary, the presence of stars is probably less than the unpaired fraction of our galaxy sample.

\section{Discussion and Conclusions}

Our principal aim was to produce the final ISW measurement using WISE, we compare our results with \cite{tomo} that used the PDR. We repeated our measurements with the WISE preliminary data and largely reproduced the individual ${C_l}^{gT}$ powers were found by \cite{tomo}, although we measured lower significance, except using alternative binning. We do not expect perfect agreement, given that the analysis was performed from the ground up.
With the same bins our best fit amplitude for the PDR was $2.5\pm1.2$ i.e. a 2.1$\sigma$ detection. The result is consistent with our 1.9$\sigma$ finding on the preliminary part of the sky but using the new data. Using the full sky we measure an ISW significance of $\sim$1.0$\sigma$.  The change is fully consistent with possible cosmic variance, as illustrated by the light gray errorbars in Figure \ref{pspec}.

With our enhanced mapmaking and better view of the WISE data we suppressed artefacts. Our mask is entirely based on the properties of the WISE object flags and many systematics were revealed. However, the signal decreased despite the improvements in our analysis methods.

While some recent studies especially \cite{tomo} raised the possibility that the ISW correlations might be higher than $\Lambda CDM$ predictions, we conclude that the signal we found is consistent with $\Lambda CDM$ and previous measurements \citep{b13,b14}. Our analysis highlighted that higher ISW amplitude measurements on certain parts of the sky can be due to cosmic variance.
\section*{Acknowledgments}
We take immense pleasure in thanking the support of NASA grants NNX12AF83G and NNX10AD53G and the Polanyi program of the Hungarian National Office for the Research and Technology (NKTH), BRG acknowledges support from PRIN INAF 2010. In addition, AK and ZF acknowledges support from OTKA through grant no. 101666. We are very thankful to Luigi Guzzo because of his help with the collaborative work of AK and BRG in Merate. Finally yet importantly, we thank the useful suggestions of the WISE team.
\bibliographystyle{mn2e}
\bibliography{refs}
\end{document}